\documentclass[graybox]{svmult}

\usepackage{mathptmx}       
\usepackage{helvet}         
\usepackage{courier}        
\usepackage{type1cm}        
\usepackage{makeidx}         
\usepackage{graphicx}        
\usepackage{multicol}        
\usepackage[bottom]{footmisc}
\newcommand{\cm}{\mathrm{c\!\:\!.m\!\:\!.}}

\makeindex             

\begin{document}

\title*{Faddeev equation approach for three-cluster nuclear reactions}
\author{A. Deltuva  \and A.~C. Fonseca \and R. Lazauskas}
\institute{A. Deltuva  \at Centro de F\'{\i}sica Nuclear da Universidade
de Lisboa, P-1649-003 Lisboa, Portugal, \email{deltuva@cii.fc.ul.pt}
\and A.~C. Fonseca \at Centro de F\'{\i}sica Nuclear da Universidade
de Lisboa, P-1649-003 Lisboa, Portugal
\and R. Lazauskas \at IPHC, IN2P3-CNRS/Universit\'e Louis Pasteur BP 28,
F-67037 Strasbourg Cedex 2, France,
\email{rimantas.lazauskas@ires.in2p3.fr}
}
%
%
\maketitle

\abstract{
In this lecture we aim to present a formalism based on Faddeev-like
equations for describing nuclear three-cluster reactions that
include elastic, transfer and breakup channels. Two different
techniques based on momentum-space and configuration-space
representations are explained in detail. An important new
feature of these methods is the possibility to account for the repulsive
Coulomb interaction between two of the three clusters in all channels.
Comparison with previous calculations based on
approximate methods used in nuclear reaction theory is also discussed.}

\section{ Introduction}

Nuclear collision experiments, performed at ion accelerators, are a very
powerful tool to study nuclear properties at low and intermediate
energies.  In order to interpret accumulated experimental data appropriate
theoretical methods are necessary enabling the simultaneous description of the
available elastic, rearrangement and breakup reactions.

Regardless of its importance, the theoretical description of quantum-mechanical
collisions turns out to be one of the most complex and slowly advancing problems
in theoretical physics. If during the last decade accurate solutions for the
nuclear bound state problem became
available, full solution of the scattering problem (containing elastic,
rearrangement and breakup channels) remains limited to the three-body case.

The main difficulty is related to the fact that, unlike the
bound state wave functions, scattering wave functions are not localized. In
configuration space one is obliged to solve  multidimensional
differential equations with extremely complex boundary conditions; by
formulating  the quantum-mechanical scattering problem in momentum space
one has to deal with  non-trivial singularities in the
kernel of  multivariable integral equations.

A rigorous mathematical formulation of the quantum mechanical three-body
problem in the framework of non relativistic dynamics has been  introduced
by Faddeev in the early sixties~\cite{Fad_60},
in the context of the three-nucleon system with short range
interactions. In momentum space these equations might be slightly
modified by formulating
them in terms of three-particle transition operators that
are smoother functions compared to the system wave functions.
Such a modification was proposed by Alt, Grassberger, and
Sandhas~\cite{alt:67a} (AGS).

Solutions of the AGS equations with short range interactions were readily  obtained in the early seventies. As large computers became available progress followed leading, by the end eighties, to fully converged solutions of these equations for neutron-deuteron ($n$-$d$)  elastic scattering and breakup using realistic short range nucleon-nucleon ($N$-$N$) interactions. Nevertheless the inclusion of the long range Coulomb force in momentum space calculations of proton-deuteron ($p$-$d$) elastic scattering and breakup with the same numerical reliability as calculations with short range interactions alone, only become possible in the last decade.


Significant progress has been achieved~\cite{deltuva:05a,deltuva:05d}
by developing the screening and renormalization
procedure for the Coulomb interaction in momentum space using a smooth
but at the same time sufficiently rapid screening. This technique permitted
to extend the calculations to the systems of three-particles with arbitrary
masses above the breakup threshold~\cite{deltuva:06b,deltuva:07d}.

However it has taken some time to formulate the
appropriate boundary conditions in configuration space for
the three-body problem~\cite{Merkuriev_71,Merkuriev_74,MGL_76}
and even longer to reformulate the original Faddeev equations to allow the
incorporation of long-range Coulomb like interactions~\cite{Merkuriev_80,Merkuriev_81}.
Rigorous solution of the three-body problem with short range interactions has
been achieved just after these theoretical developments, both below and above
breakup threshold. On the other hand the numerical solution for the three-body
problem  including charged particles above
the three-particle breakup threshold  has  been achieved only recently.
First it has been done by using approximate Merkuriev
boundary conditions in configuration space~\cite{kievsky:97}.
Nevertheless this approach proved to be a rather complex task numerically,
remaining unexplored beyond the $p$-$d$ scattering case, but not yet for the $p$-$d$ breakup.

Finally, very recently configuration space method based on complex scaling have been
developed and applied for $p$-$d$ scattering~\cite{lazauskas:11a}.
This method allows to treat
the scattering problem  using very simple boundary conditions, equivalent
to the ones employed to solve the bound-state problem.

\bigskip
The aim of this lecture is to present these two recently developed
techniques, namely the momentum-space method based on screening and renormalization
as well as the configuration-space complex scaling method.
This lecture is structured as follows: the first part serves to
introduce theoretical formalisms for momentum space and configuration
space calculations;
in the second part we present some selected calculations with an
aim to test the performance and validity of the two presented methods.

\section{Momentum-space description of three-particle scattering}
\label{sec:p}

We describe the scattering process in a system of three-particles
interacting via pairwise short-range potentials
 $v_\alpha$, $\alpha=1,2,3$; we use the odd-man-out notation,
that is, $v_1$ is the potential between particles 2 and 3.
In the framework of nonrelativistic quantum mechanics the center-of-mass
(c.m.) and the internal motion can be separated by introducing
Jacobi momenta
\begin{eqnarray}\label{eq:Jacobi}
    \vec{p}_\alpha & = &\frac{m_{\gamma} \vec{k}_\beta - m_{\beta} \vec{k}_\gamma }
        {m_{\beta} + m_{\gamma} },  \\
    \vec{q}_\alpha & = &
    \frac{m_{\alpha} (\vec{k}_\beta + \vec{k}_\gamma) -
      (m_{\beta} + m_{\gamma}) \vec{k}_\alpha }
         {m_{\alpha} + m_{\beta} + m_{\gamma} }, 
\end{eqnarray}
 with ($\alpha \beta \gamma $) being cyclic permutations of (123);
$\vec{k}_\alpha$ and $m_{\alpha}$ are the individual particle
momenta and masses, respectively. The c.m. motion is free and in the
following we consider only the internal motion; the corresponding
kinetic energy operator is $H_0$ while the full Hamiltonian is
\begin{equation} \label{eq:H}
 H = H_0 + \sum_{\alpha=1}^3   v_\alpha .
\end{equation}

\subsection{Alt, Grassberger, and Sandhas equations}

We consider the  particle $\alpha$ scattering from the pair  $\alpha$
that is bound with energy $ \epsilon_\alpha$.
The initial channel state $|b_{\alpha}\vec{q}_\alpha\rangle$
is the product of the bound state wave function $|b_\alpha \rangle$
for the pair $\alpha$ and a plane wave with the
relative particle-pair  $\alpha$ momentum $\mathbf{q}_\alpha$;
the dependence on the discrete quantum numbers is suppressed
in our notation. $|b_{\alpha}\vec{q}_\alpha\rangle$
is the eigenstate of the corresponding channel Hamiltonian
$H_\alpha = H_0 + v_\alpha$
with the energy eigenvalue $E= \epsilon_\alpha + q^2_\alpha/2M_\alpha$
where 
$M_\alpha$ is the particle-pair  $\alpha$ reduced mass.
 The final channel state is the particle-pair state in the same or
different configuration $|b_{\beta}\vec{q}_\beta\rangle$
in the case of elastic and rearrangement scattering
or, in the case of breakup,
 it is the  state of three free particles
 $|\vec{p}_{\gamma}\vec{q}_\gamma\rangle$ with the same
energy $E= p_\gamma^2/2\mu_\gamma +  q_\gamma^2/2M_\gamma $ and pair $\gamma$
reduced mass $\mu_\gamma$; 
any set of Jacobi momenta can be used equally well for the breakup state.

The stationary scattering states~\cite{schmid:74a,gloeckle:83a}
corresponding to the above channel states are eigenstates of the full
Hamiltonian; they are obtained from the channel states using
the full resolvent $G = (E+i0-H)^{-1}$, i.e.,
\begin{eqnarray} \label{eq:psi_a}
  |b_\alpha  \vec{q}_\alpha \rangle^{(+)} & = &
   i0  G |b_\alpha \vec{q}_\alpha \rangle, \\
\label{eq:psi_0}
  |\vec{p}_\alpha\vec{q}_\alpha \rangle^{(+)} & = &
   i0 G |\vec{p}_\alpha\vec{q}_\alpha \rangle.
\end{eqnarray}
The full resolvent $G$ may be decomposed into the channel resolvents
$G_\beta  = (E+i0-H_\beta)^{-1}$ and/or free resolvent $G_0  = (E+i0-H_0)^{-1}$
as
\begin{equation}
G = G_\beta  + G_\beta   \bar{v}_\beta G ,
\end{equation}
with $\beta=0,1,2,3$ and
$ \bar{v}_\beta = \sum_{\gamma=1}^3 \bar{\delta}_{\beta \gamma} v_\gamma$
where $\bar{\delta}_{\beta \gamma} = 1-{\delta}_{\beta \gamma}$.
Furthermore, the channel resolvents
\begin{equation}
G_\beta  = G_0 + G_0   T_\beta G_0 ,
\end{equation}
can be related to the corresponding two-particle transition operators
\begin{equation}
  T_\beta = v_\beta + v_\beta G_0 T_\beta ,
\end{equation}
embedded into three-particle Hilbert space. Using these definitions
Eqs.~(\ref{eq:psi_a}) and (\ref{eq:psi_0})
can be written as  triads of Lippmann-Schwinger equations
\begin{eqnarray} \label{eq:psi_LS}
|b_\alpha  \vec{q}_\alpha \rangle^{(+)} & =  {} &
  \delta_{\beta \alpha}  |b_\alpha \vec{q}_\alpha \rangle
 + G_\beta  \bar{v}_\beta |b_\alpha  \vec{q}_\alpha \rangle^{(+)} , \\
 |\vec{p}_\alpha\vec{q}_\alpha \rangle^{(+)} & =  {} &
 (1+ G_0  T_\beta ) |\vec{p}_\alpha\vec{q}_\alpha \rangle
 + G_\beta \bar{v}_\beta  |\vec{p}_\alpha\vec{q}_\alpha \rangle^{(+)} ,
\end{eqnarray}
with $\alpha$ being fixed and $\beta =1,2,3$; they are necessary and sufficient
to define the states $|b_\alpha  \vec{q}_\alpha \rangle^{(+)}$ and
$|\vec{p}_\alpha\vec{q}_\alpha \rangle^{(+)}$ uniquely.
However, in scattering problems it may be more convenient to work with
the multichannel transition operators $U_{\beta \alpha}$ defined such that
their on-shell elements yield scattering amplitudes, i.e.,
\begin{equation} \label{eq:U-V}
U_{\beta \alpha} |b_\alpha \vec{q}_\alpha \rangle =
 \bar{v}_\beta |b_\alpha  \vec{q}_\alpha \rangle^{(+)}.
\end{equation}
 Our calculations  are based on the AGS version~\cite{alt:67a} of
three-particle scattering theory. In accordance with Eq.~(\ref{eq:U-V})
it defines the multichannel transition operators $U_{\beta \alpha}$
by the decomposition of the full resolvent $ G$ into channel
and/or free resolvents as
\begin{equation} \label{eq:G-U}
G = \delta_{\beta \alpha} G_\alpha  + G_\beta U_{\beta \alpha} G_\alpha .
\end{equation}
The multichannel transition operators $U_{\beta \alpha}$
with fixed $\alpha$ and $\beta = 1,2,3$  are solutions
of three coupled integral equations
\begin{equation} \label{eq:AGSnsym_a}
  U_{\beta \alpha} =  \bar{\delta}_{\beta \alpha} G_0^{-1} +
  \sum_{\gamma=1}^3 \bar{\delta}_{\beta \gamma} T_{\gamma} G_0 U_{\gamma \alpha}.
\end{equation}
The transition matrix $U_{0 \alpha} $ to final states with three free
particles can be obtained from the solutions of Eq.~(\ref{eq:AGSnsym_a})
by quadrature, i.e.,
\begin{equation} \label{eq:AGSnsym_b}
  U_{0 \alpha} = G_0^{-1} + \sum_{\gamma=1}^3 T_{\gamma} G_0 U_{\gamma \alpha}.
\end{equation}

The on-shell matrix elements
$\langle b_{\beta} \vec{q}'_\beta |U_{\beta \alpha} |b_\alpha \vec{q}_\alpha \rangle$
are  amplitudes (up to a factor) for  elastic
($\beta = \alpha$) and rearrangement ($\beta \neq \alpha$) scattering.
For example, the differential cross section for the
$\alpha + (\beta\gamma) \to \beta + (\gamma\alpha)$ reaction in the c.m.
system is given by
\begin{equation} \label{eq:dcsab}
\frac{d \sigma_{\alpha \to \beta}}{d \Omega_\beta} =
(2\pi)^4 M_\alpha M_\beta \frac{q'_\beta}{q_\alpha}
| \langle b_{\beta} \vec{q}'_\beta |U_{\beta \alpha}
|b_\alpha \vec{q}_\alpha \rangle|^2.
\end{equation}
The cross section for the breakup is determined by the on-shell matrix elements
$\langle \vec{p}'_{\gamma} \vec{q}'_\gamma |U_{0 \alpha}
|b_\alpha \vec{q}_\alpha \rangle$.
Thus, in the AGS framework all elastic, rearrangement, and breakup reactions
are calculated on the same footing.

Finally we note that the AGS equations can be extended to include
also the three-body forces as done in Ref.~\cite{deltuva:09e}.

\subsection{Inclusion of the Coulomb interaction}

The Coulomb potential $w_C$, due to its long range, does not satisfy the
mathematical properties required for the formulation of standard scattering theory as given in the previous subsection
for short-range interactions $v_\alpha$. However,
in nature the Coulomb potential is always screened at large distances.
The comparison of the data from typical nuclear physics experiments
and  theoretical  predictions with full Coulomb  is meaningful
only if the full and screened Coulomb become physically indistinguishable.
This was proved in Refs.~\cite{taylor:74a,semon:75a}
where the screening and renormalization method for the scattering of two
charged particles was proposed.
We base our treatment of the Coulomb interaction  on that idea.

Although we use momentum-space framework, we first choose the
screened Coulomb potential in configuration-space representation as
\begin{equation} \label{eq:wr}
w_R(r) = w_C(r)\; e^{-(r/R)^n} ,
\end{equation}
and then transform it to momentum-space.
Here $R$ is the screening radius
and $n$ controls the smoothness of the screening.
The standard scattering theory is formally applicable to the screened
Coulomb potential $w_R$, i.e.,  the Lippmann-Schwinger equation yields
the two-particle transition matrix
\begin{equation} \label{eq:tr}
t_R = w_R + w_R g_0 t_R ,
\end{equation}
where $g_0$ is the two-particle free resolvent.
It was proven in Ref.~\cite{taylor:74a}  that
in the limit of infinite screening radius $R$ the on-shell screened Coulomb
transition matrix (screened Coulomb scattering amplitude)
$\langle \mathbf{p}'| t_R |  \mathbf{p} \rangle$ with $p'=p$,
renormalized by an infinitely oscillating phase factor
$z_R^{-1}(p) = e^{2i\phi_R(p)}$, approaches the  full Coulomb amplitude
 $\langle \mathbf{p}'| t_C |  \mathbf{p} \rangle$
in general as a distribution.
The convergence in the sense of distributions
is sufficient for the description of physical observables in a real
experiment where the incoming beam is not a plane wave but wave packet
and therefore   the cross section
is determined not directly by the scattering amplitude but
by the outgoing wave packet, i.e., by the scattering amplitude averaged
over the initial state physical wave packet.
In practical calculations~\cite{alt:02a,deltuva:05a} this averaging
is carried out implicitly, replacing the renormalized screened Coulomb
amplitude in the $R \to \infty$ limit by the full one, i.e.,
\begin{equation} \label{eq:taylor2}
\lim_{R \to \infty} z_R^{-1}(p)
 \langle \mathbf{p}'| t_R |  \mathbf{p} \rangle \to
 \langle \mathbf{p}'| t_C |  \mathbf{p} \rangle.
\end{equation}
Since $z_R^{-1}(p)$ is only a phase factor,
the above relations indeed demonstrate that the physical observables
become insensitive to screening provided it takes place at sufficiently
large distances $R$ and, in the $R \to \infty$ limit,
coincide with the corresponding quantities referring to the full Coulomb.
Furthermore,
renormalization by $ z_{R}^{-\frac12}(p_i)$ in the $R \to \infty$ limit relates
also the screened and full Coulomb wave functions~\cite{gorshkov:61}, i.e.,
\begin{equation} \label{eq:gorshkov}
\lim_{R \to \infty} (1 + g_0 t_R) |\mathbf{p} \rangle z_R^{-\frac12}(p)
    =  |\psi_C^{(+)}(\mathbf{p}) \rangle.
\end{equation}

The screening and renormalization method based on the above relations
can be extended to more complicated systems, albeit with some limitations.
We consider the system of  three-particles  with charges $z_\alpha$
of equal sign interacting via pairwise strong short-range and screened
Coulomb potentials $v_\alpha + w_{\alpha R}$ with $\alpha$ being 1, 2, or 3.
The corresponding two-particle transition matrices are calculated with
the full channel interaction
\begin{equation} \label{eq:TR}
   T^{(R)}_\alpha  =  (v_\alpha + w_{\alpha R}) +
     (v_\alpha + w_{\alpha R})  G_0 T^{(R)}_\alpha,
  \end{equation}
and the multichannel transition operators $U^{(R)}_{\beta \alpha}$
for elastic and rearrangement scattering are solutions
of the AGS equation
\begin{equation}
 U^{(R)}_{\beta \alpha}  =  \bar{\delta}_{\beta \alpha} G_0^{-1}
+ \sum_{\gamma=1}^3  \bar{\delta}_{\beta \gamma}
T^{(R)}_\gamma G_0 U^{(R)}_{\gamma \alpha} ;
\label{eq:Uba}
\end{equation}
all operators depend parametrically on the Coulomb screening radius $R$.

In order to isolate the screened Coulomb contributions to the transition
amplitude that diverge in the infinite $R$ limit
 we introduce an auxiliary screened Coulomb potential $W^{\cm}_{\alpha R}$
between the  particle $\alpha$
and the center of mass (c.m.) of the remaining pair.
The same screening function has to be used for both Coulomb potentials
$w_{\alpha R}$ and $W^{\cm}_{\alpha R}$.
The corresponding transition matrix
\begin{equation} \label{eq:Tcm}
T^{\cm}_{\alpha R}  = W^{\cm}_{\alpha R} +
W^{\cm}_{\alpha R} G^{(R)}_{\alpha}  T^{\cm}_{\alpha R} ,
\end{equation}
with $ G^{(R)}_{\alpha} = (E+i0-H_0-v_\alpha - w_{\alpha R})^{-1}$
is a two-body-like operator and therefore its on-shell and half-shell
behavior in the limit $R \to \infty$
is given by Eqs.~(\ref{eq:taylor2}) and (\ref{eq:gorshkov}).
As derived in Ref.~\cite{deltuva:05a}, the
three-particle transition operators may be decomposed as
\begin{eqnarray}
    U^{(R)}_{\beta \alpha} &=& \delta_{\beta\alpha} T^{\cm}_{\alpha R}
    +  [1 + T^{\cm}_{\beta R} G^{(R)}_{\beta}] 
    \tilde{U}^{(R)}_{\beta\alpha}  
          [1 + G^{(R)}_{\alpha} T^{\cm}_{\alpha R}] \quad
\label{eq:U-T} \\
 &=& \delta_{\beta\alpha} T^{\cm}_{\alpha R}
+ (U^{(R)}_{\beta \alpha} - \delta_{\beta\alpha} T^{\cm}_{\alpha R}).
\label{eq:U-T2}
\end{eqnarray}
where the auxiliary operator $\tilde{U}^{(R)}_{\beta\alpha}$ is of
short range when calculated between on-shell screened Coulomb states.
Thus, the three-particle transition operator $U^{(R)}_{\beta \alpha}$
has a long-range part $\delta_{\beta\alpha} T^{\cm}_{\alpha R}$ whereas the
remainder  $U^{(R)}_{\beta \alpha} - \delta_{\beta\alpha} T^{\cm}_{\alpha R}$
is a short-range operator that  is externally distorted
due to the screened Coulomb waves generated by
$[1 + G^{(R)}_{\alpha} T^{\cm}_{\alpha R}]$.
On-shell, both parts do not have a proper limit as $R \to \infty$ but
the limit exists after renormalization by an appropriate phase factor,
yielding  the transition amplitude for full Coulomb
\begin{eqnarray} \nonumber
&&    \langle  b_\beta \mathbf{q}'_\beta | U^{(C)}_{\beta \alpha}
    |b_\alpha \mathbf{q}_\alpha\rangle     =
    \delta_{\beta \alpha}
    \langle b_\alpha \mathbf{q}'_\beta |T^{\cm}_{\alpha C}
    |b_\alpha \mathbf{q}_\alpha \rangle  \\ & &
 +   \lim_{R \to \infty} [  Z^{-\frac{1}{2}}_{\beta R}(q'_\beta)
    \langle b_\beta \mathbf{q}'_\beta  |
            ( U^{(R)}_{\beta \alpha}  
             - \delta_{\beta\alpha} T^{\cm}_{\alpha R})
            |b_\alpha \mathbf{q}_\alpha \rangle
             Z^{-\frac{1}{2}}_{\alpha R}(q_\alpha) ]. \quad
\label{eq:UC2}
\end{eqnarray}
The first term on the right-hand side of Eq.~(\ref{eq:UC2}) is known
analytically~\cite{taylor:74a}; it corresponds to the  particle-pair $\alpha$
full Coulomb transition amplitude
that results from the implicit renormalization of $T^{\cm}_{\alpha R}$
according to Eq.~(\ref{eq:taylor2}).
The $R \to \infty$ limit for the remaining part
$( U^{(R)}_{\beta \alpha} - \delta_{\beta\alpha} T^{\cm}_{\alpha R})$
of the multichannel transition matrix is performed numerically;
due to the short-range nature of this term
the convergence with the increasing screening radius $R$
is fast and the limit is reached with sufficient accuracy at
finite $R$; furthermore, it can be calculated using the partial-wave expansion.
We emphasize that  Eq.~(\ref{eq:UC2}) is by no means an approximation
since it is based on the obviously exact identity (\ref{eq:U-T2})
where the  $R \to \infty$ limit for each term exists and
is calculated separately.

The renormalization factor for  $R \to \infty $ is a diverging phase factor
  \begin{equation}
    Z_{\alpha R}(q_\alpha) = e^{-2i \Phi_{\alpha R}(q_\alpha)},
  \end{equation}
  where $\Phi_{\alpha R}(q_\alpha)$, though independent of the particle-pair
relative angular momentum $l_\alpha$ in the infinite $R$ limit,
may be realized by
  \begin{equation}    \label{eq:phiRl}
\Phi_{\alpha R}(q_\alpha) = \sigma_{l_\alpha}^{\alpha}(q_\alpha) -
\eta_{l_\alpha R}^{\alpha}(q_\alpha),
  \end{equation}
with the diverging screened Coulomb phase shift
$\eta_{l_\alpha R}^{\alpha}(q_\alpha)$
corresponding to standard boundary conditions
and the proper Coulomb one $\sigma_{l_\alpha}^{\alpha}(q_\alpha)$ referring to the
logarithmically distorted proper Coulomb boundary conditions.
For the screened Coulomb potential of Eq.~(\ref{eq:wr})
the infinite $R$ limit of $\Phi_{\alpha R}(q_\alpha)$ is known analytically,
\begin{equation}  \label{eq:phiRlln}
\Phi_{\alpha R}(q_\alpha)=\mathcal{K}_{\alpha}(q_\alpha)[\ln{(2q_\alpha R)} - C/n],
  \end{equation}
where  $C \approx 0.5772156649$ is the Euler number and
$\mathcal{K}_{\alpha}(q_\alpha) = \alpha_{e.m.}z_\alpha \sum_\gamma
\bar{\delta}_{\gamma\alpha} z_\gamma M_\alpha/q_\alpha$
is the Coulomb parameter with $\alpha_{e.m.} \approx 1/137$.
The form of the renormalization phase $\Phi_{\alpha R}(q_\alpha)$ to be used
in the actual calculations with finite screening radii $R$ is not unique,
but the converged results show independence of
the chosen form of $\Phi_{\alpha R}(q_\alpha)$.

For breakup reactions we follow a  similar strategy. However,
the proper three-body Coulomb wave function and its relation to the
three-body screened Coulomb wave function  is, in general, unknown.
This prevents the application of the screening and renormalization method to the
reactions involving three free charged particles (nucleons or nuclei)
in the final state.
However, in  the system of two charged particles and a neutral one
with $z_\rho = 0$, the final-state Coulomb distortion becomes again
a two-body problem with the screened Coulomb transition matrix
\begin{equation}
  T_{\rho R} =  w_{\rho R} + w_{\rho R} G_0 T_{\rho R}.
\end{equation}
This makes the channel $\rho$, corresponding to the correlated pair
of charged particles,  the most convenient choice for the
description of the final breakup state.
As shown in Ref.~\cite{deltuva:05d}, the  AGS
breakup operator
 \begin{equation}\label{eq:U0a}
      U^{(R)}_{0\alpha} = {}
G_0^{-1} + \sum_{\gamma=1}^3 T^{(R)}_{\gamma} G_0 U^{(R)}_{\gamma \alpha} ,
  \end{equation}
can be decomposed as
  \begin{equation}\label{eq:U0t}
      U^{(R)}_{0\alpha} = {}  (1 + T_{\rho R} G_{0})
      \tilde{U}^{(R)}_{0\alpha} (1 + G^{(R)}_{\alpha} T^{\cm}_{\alpha R}),
  \end{equation}
where the reduced operator
$\tilde{U}^{(R)}_{0\alpha}(Z)$ calculated between screened Coulomb
distorted initial and final states is of finite range.
In the full breakup operator $U^{(R)}_{0 \alpha}(Z)$
the external distortions show up in screened
Coulomb waves generated by $(1 + G^{(R)}_{\alpha} T^{\cm}_{\alpha R})$
in the initial state and by $(1 + T_{\rho R} G_{0})$ in the final
state; both wave functions do not have proper limits as $R \to \infty$.
Therefore  the full breakup transition amplitude  in the case of
the unscreened Coulomb potential is obtained via the renormalization
of the on-shell breakup transition matrix $ U^{(R)}_{0 \alpha}$
in the infinite $R$ limit
\begin{equation}
      \langle \mathbf{p}'_\rho \mathbf{q}'_\rho  |  U^{(C)}_{0 \alpha}
      |b_\alpha \mathbf{q}_\alpha  \rangle  =
      \lim_{R \to \infty} [ z^{-\frac{1}{2}}_{\rho R}(p'_\rho)
      \langle \mathbf{p}'_\rho \mathbf{q}'_\rho | U^{(R)}_{0 \alpha}
      |b_\alpha \mathbf{q}_\alpha \rangle Z_{\alpha R}^{-\frac{1}{2}}(q_\alpha )],
\label{eq:UC1a}
 \end{equation}
where  $\mathbf{p}'_\rho$ is the relative momentum between the charged
particles in the  final state,
$\mathbf{q}'_\rho$ the corresponding particle-pair  relative momentum, and
\begin{equation}  \label{eq:phiRp}
z_{\rho R}(p'_\rho) = e^{-2i\kappa_\rho(p'_\rho)[\ln{(2p'_\rho R)} - C/n]} ,
  \end{equation}
the final-state renormalization factor with the Coulomb parameter
$\kappa_\rho(p'_\rho)$ for the pair $\rho$.
The limit in Eq.~(\ref{eq:UC1a}) has to be performed numerically,
but, due to the short-range nature of the breakup operator,
the convergence with increasing screening radius $R$
is fast and the limit is reached with sufficient accuracy at
finite $R$.
Thus, to include the Coulomb interaction via the screening and renormalization
method one only needs to solve standard scattering theory equations.

\subsection{Practical realization}

\begin{figure}[t]
\sidecaption[t]
\includegraphics[scale=.55]{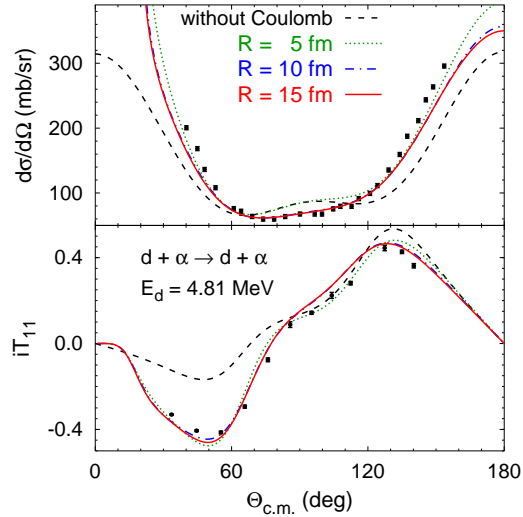}
\caption{
Differential cross section and deuteron vector analyzing power $iT_{11}$
of the $\alpha d$ elastic scattering  at 4.81~MeV deuteron lab energy
as functions of the c.m. scattering angle.
Convergence with the screening radius $R$ used to calculate the
short-range part of the amplitudes is studied:
 $R= 5$~fm (dotted curves),  $R= 10$~fm (dash-dotted curves), and
 $R= 15$~fm (solid curves). Results without Coulomb
are given by dashed curves.
The experimental data  are from Refs.~\cite{bruno:80,gruebler:70a}. }
\label{fig:Rad}
\end{figure}

\begin{figure}[t]
\sidecaption[t]
\includegraphics[scale=.55]{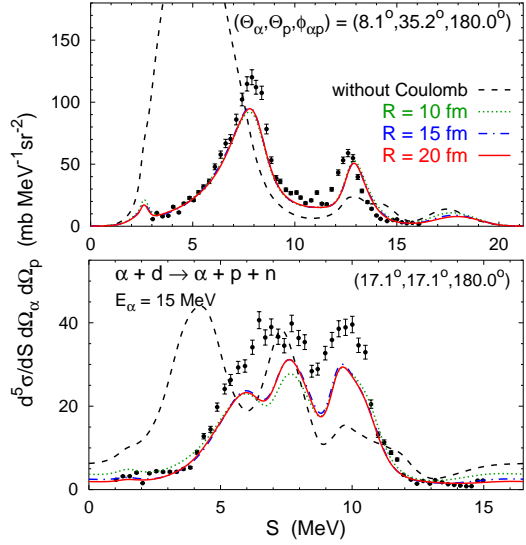}
\caption{
Fivefold differential cross section of the $\alpha d$ breakup  reaction
at 15~MeV $\alpha$ lab energy for several combinations of
 $\alpha$ and proton scattering angles as function of the final-state
energy variable $S$ with $dS = (dE_\alpha^2 + dE_p^2)^{1/2}$.
Convergence with the screening radius $R$ is studied:
 $R= 10$~fm (dotted curves),  $R= 15$~fm (dash-dotted curves), and
 $R= 20$~fm (solid curves). Results without Coulomb
are given by dashed curves.
The experimental data  are from Ref.~\cite{koersner:77}. }
\label{fig:Radb}
\end{figure}

We calculate the short-range part of the elastic, rearrangement, and breakup
scattering amplitudes (\ref{eq:UC2}) and (\ref{eq:UC1a}) by
solving standard scattering  equations (\ref{eq:Uba}), (\ref{eq:Tcm}),
 and (\ref{eq:U0a}) with a finite  Coulomb screening radius $R$.
We work in the momentum-space partial-wave basis~\cite{deltuva:phd},
i.e., we use three sets \\
$|p_\alpha q_\alpha \nu_\alpha \rangle \equiv
|p_\alpha q_\alpha (l_\alpha \{ [L_\alpha(s_\beta s_\gamma)S_\alpha]
I_\alpha s_\alpha \} K_\alpha) { J M} \rangle$
with $(\alpha,\beta,\gamma)$ being cyclic permutations of (1,2,3).
Here $s_\alpha$ is the spin of particle $\alpha$, $L_\alpha$ and $l_\alpha$ are
the orbital angular momenta associated with $p_\alpha$ and  $q_\alpha$ respectively,
whereas $S_\alpha$, $I_\alpha$, and $K_\alpha$ are intermediate angular momenta
that are coupled to a total angular momentum $J$ with projection $M$.
All discrete quantum numbers are abbreviated by $\nu_\alpha$.
The integration over the momentum variables is discretized using
Gaussian quadrature rules thereby converting a system of integral equations
for each $J$ and parity $\Pi = (-)^{L_\alpha +l_\alpha}$
into a very large system of linear algebraic equations.
Due to the huge dimension those linear systems cannot be solved directly.
Instead we expand the AGS transition operators (\ref{eq:Uba})
into the corresponding Neumann series
\begin{equation} \label{eq:neumann}
 U^{(R)}_{\beta \alpha} = \bar{\delta}_{ \beta  \alpha }  G_0^{-1} +
\sum_{\gamma=1}^3 \bar{\delta}_{ \beta  \gamma } T^{(R)}_\gamma
\bar{\delta}_{ \gamma  \alpha }
 + \sum_{\gamma=1}^3 \bar{\delta}_{ \beta  \gamma } T^{(R)}_\gamma G_0
\sum_{\sigma=1}^3  \bar{\delta}_{\gamma \sigma }  T^{(R)}_\sigma
\bar{\delta}_{\sigma  \alpha}
   + \cdots  ,
\end{equation}
that are summed up by the iterative Pade method~\cite{chmielewski:03a};
it yields an accurate solution of Eq.~(\ref{eq:Uba}) even when the
Neumann series (\ref{eq:neumann}) diverges.
Each two-particle transition operator $ T^{(R)}_{\gamma}$ is evaluated
in its proper basis $|p_\gamma q_\gamma \nu_\gamma \rangle$, thus,
transformations between all three bases are needed.
The calculation of the involved overlap functions
$ \langle p_\beta q_\beta \nu_\beta |p_\alpha q_\alpha \nu_\alpha \rangle$
follows closely the calculation of three-nucleon permutation
operators discussed in Refs.~\cite{deltuva:phd,gloeckle:83a}.
A special treatment~\cite{chmielewski:03a,deltuva:phd}
is needed for the integrable singularities
arising from the pair bound state poles in $ T^{(R)}_{\gamma}$  and from $G_0$.
Furthermore, we have to make sure that $R$ is large enough to achieve
(after renormalization) the $R$-independence of the results up to a desired
accuracy. However, those $R$ values are larger
than the range of the nuclear interaction resulting in a slower convergence
of the partial-wave expansion. As we found in Ref.~\cite{deltuva:05a},
the practical success of the screening and renormalization method depends
very much on the choice of the screening function, in our case on the power
$n$ in Eq.~(\ref{eq:wr}). We want to ensure that the
screened Coulomb potential $w_R$ approximates well the true Coulomb one
$w_C$ for distances $r<R$  and simultaneously vanishes rapidly for $r>R$,
providing a comparatively fast convergence of the partial-wave expansion.
As shown in Ref.~\cite{deltuva:05a}, this is not the case for simple
exponential screening  $(n =1)$ whereas the sharp cutoff  $(n \to \infty)$
yields slow oscillating convergence with the screening radius $R$. However,
we found that values of $3 \le n \le 8$ provide a sufficiently smooth and
rapid screening around $r=R$. The screening functions
for different $n$ values are compared in  Ref.~\cite{deltuva:05a}
together with the results demonstrating the superiority of our
optimal choice: using $3 \le n \le 8$ the convergence with the
screening radius $R$, at which the short range part of the amplitudes was
calculated, is fast enough such that the convergence of the
partial-wave expansion, though being slower than for the nuclear interaction
alone, can be achieved and there is no need to work in a plane-wave basis.
Here we use $n=4$ and show in Figs.~\ref{fig:Rad}and \ref{fig:Radb}
few examples for the $R$-convergence of the $\alpha$-deuteron scattering
observables calculated in a three-body model $(\alpha,p,n)$;
the nuclear interaction is taken from Ref.~\cite{deltuva:06b}.
The convergence with $R$ is impressively fast for both
 $\alpha$-deuteron elastic scattering and breakup. In addition we note
that the Coulomb effect is very large and clearly improves the description
of the experimental data, especially for the
differential cross section in $\alpha$-deuteron breakup reaction.
This is due to the shift of
the  $\alpha p$ $P$-wave resonance position when the $\alpha p$ Coulomb
repulsion is included that leads to
the corresponding changes in the structure of the observables.

In addition to the internal reliability criterion of the screening and
renormalization method --- the convergence with $R$ --- we note that our
results for proton-deuteron elastic scattering~\cite{deltuva:05b}
agree well over a
broad energy range with those of Ref.~\cite{kievsky:01a} obtained from
the variational configuration-space solution
of the three-nucleon Schr\"odinger equation with  unscreened Coulomb potential
 and imposing  the proper Coulomb boundary conditions explicitly.

\section{Configuration space}\label{sec:r}

In contrast to the momentum-space representation,
the  Coulomb interaction  has a trivial expression
 in configuration space and thus may seem to be easier to handle. However the
major obstacle for configuration-space treatment of the scattering problem
is related with the complexity of the wave function asymptotic structure,
which strongly complicates once three-particle breakup is available. Although for
short range interactions the analytical behavior of the breakup asymptote
of the configuration space wave function is well established, this is not a
case once long range interactions (like Coulomb) are present. Therefore a
method which enables the scattering problem to be solved without explicit use
of the wave function asymptotic form is of great importance. The
complex scaling method has been proposed~\cite{Nuttal_csm,CSM_71} and successfully
applied to calculate the resonance positions~\cite{Moiseyev} by
using bound state  boundary conditions.
As has been demonstrated recently this method can be extended also for the
scattering problem~\cite{CSM_Curdy_04,Elander_CSM}.  We
demonstrate here that this method may be also
 successfully applied to solve
three-particle scattering problems which include the long-range Coulomb interaction together
with short range optical potentials.

\subsection{Faddeev-Merkuriev equations}

Like in the momentum space formalism described above Jacobi coordinates are
also used in configuration space to separate the
center of mass of the three-particle system. One has three equivalent sets
of three-particle Jacobi coordinates
\begin{eqnarray}
\mathbf{x}_{\alpha } &=&\sqrt{\frac{2m_{\beta }m_{\gamma }}{(m_{\beta
}+m_{\gamma })m}}(\mathbf{r}_{\gamma }-\mathbf{r}_{\beta }) , \\
\mathbf{y}_{\alpha } &=&\sqrt{\frac{2m_{\beta }(m_{\beta }+m_{\gamma })}{%
(m_{\alpha }+m_{\beta }+m_{\gamma })m}}(\mathbf{r}_{\alpha }-\frac{m_{\beta }%
\mathbf{r}_{\beta }+m_{\gamma }\mathbf{r}_{\gamma }}{m_{\beta }+m_{\gamma }}) ,
\nonumber
\end{eqnarray}%
here $r_{\alpha }$ and $m_{\alpha }$ are individual particle position
vectors and masses, respectively. The choice of a mass scale $m$ is
arbitrary. The three-particle problem is formulated here using Faddeev-Merkuriev
(FM) equations~\cite{Merkuriev_80}:
\begin{eqnarray}
(E-H_{0}-\sum_{\kappa =1}^{3}w_{i}^{l})\psi _{\alpha }=(v_{\alpha
}+w_{\alpha }^{s})(\psi _{\alpha }+\psi _{\beta }+\psi _{\gamma })\nonumber , \\
(E-H_{0}-\sum_{\kappa =1}^{3}w_{i}^{l})\psi _{\beta }=(v_{\beta }+w_{\beta
}^{s})(\psi _{\alpha }+\psi _{\beta }+\psi _{\gamma }) , \\
(E-H_{0}-\sum_{\kappa =1}^{3}w_{i}^{l})\psi _{\gamma }=(v_{\gamma
}+w_{\gamma }^{s})(\psi _{\alpha }+\psi _{\beta }+\psi _{\gamma }) , \nonumber%
\end{eqnarray}%
where the Coulomb interaction is split in two parts (short and long range), $%
w_{\alpha }=w_{\alpha }^{s}+w_{\alpha }^{l}$, by means of some arbitrary
cut-off function $\chi _{\alpha }(x_{\alpha },y_{\alpha })$:
\begin{equation}
w_{\alpha }^{s}(x_{\alpha },y_{\alpha })=w_{\alpha }(x_{\alpha })\chi
_{\alpha }(x_{\alpha },y_{\alpha })\qquad w_{\alpha }^{l}(x_{\alpha
},y_{\alpha })=w_{\alpha }(x_{\alpha })[1-\chi _{\alpha }(x_{\alpha
},y_{\alpha })]
\end{equation}%
This cut-off function intends to shift the full Coulomb interaction
in the $w_{\alpha }^{s}$ term if $%
x_{\alpha }$ is small, whereas the $w_{\alpha }^{l}$ term acquires the full Coulomb
interaction if $x_{\alpha }$ becomes large and $y_{\alpha }<x_{\alpha }$.
The practical choice of function $\chi _{\alpha }(x_{\alpha },y_{\alpha })$
has been proposed in~\cite{Merkuriev_80}:
\begin{equation}
\chi _{\alpha }(x_{\alpha },y_{\alpha })=\frac{2}{[1+exp{(\frac{[x_{\alpha }/x_0]^\mu}{1+y_{\alpha }/y_0})}]} ,
\end{equation}%
with free parameters $x_{0},y_{0}$ having size comparable with the
charge radii of the respective binary systems; the value of parameter
$\mu$ must be larger than 1 and is usually set $\mu\approx2$.
 In such a way the so-called
Faddeev amplitude $\psi _{\alpha }$ intends to acquire full asymptotic
behavior of the binary $\alpha -(\beta \gamma )$ channels, i.e:
\begin{eqnarray}
\psi _{\alpha }(\mathbf{x}_{\alpha },\mathbf{y}_{\alpha }\rightarrow \infty
)=\delta _{\kappa ,\alpha }\psi _{\alpha }^{i_{\kappa }}(\mathbf{x}_{\alpha
})\phi _{\alpha }^{i_{\kappa },in}(\mathbf{y}_{\alpha })&+&\sum_{j_{\alpha
}}f_{j_{\alpha i_{\kappa }}}(\mathbf{x}_{\alpha }.\mathbf{y}_{\alpha })\psi
_{\alpha }^{j_{\alpha }}(\mathbf{x}_{\alpha })\phi _{\alpha }^{j_{\alpha
},out}(\mathbf{y}_{\alpha })\nonumber \\
&+&A_{i_{\kappa }}(\mathbf{x}_{\alpha },\mathbf{y}%
_{\alpha })\Phi _{i_{\kappa }}^{out}(\mathbf{\rho }) ,
\end{eqnarray}
where the hyperradius is $\rho =\sqrt{x_{\alpha }^{2}+y_{\alpha }^{2}}$. An
expression $\varphi _{\alpha }^{i_{\alpha }}(\mathbf{x}_{\alpha })\phi
_{\alpha }^{i_{\kappa },in}(\mathbf{y}_{\alpha })$ represents the incoming wave for particle $\alpha $ on pair $(\beta \gamma )$ in
the bound state $i_{\alpha }$, with $\varphi _{\alpha }^{i_{\alpha }}(%
\mathbf{x}_{\alpha })$ representing the normalized wave
function of bound state $i_{\alpha }$. This  wave
function is a solution of the $(E-H_{0}-w_{\alpha }-v_{\alpha
}-W_{\alpha }^{c.m.})$ two-body Hamiltonian. The $\phi _{\alpha }^{j_{\alpha },out}(%
\mathbf{y}_{\alpha })$ and $\Phi _{i_{\kappa }}^{out}(\mathbf{\rho }_{\alpha
})$ represent outgoing waves for binary and three-particle breakup channels
respectively. In the asymptote, one has the following behavior:
\begin{eqnarray}
\varphi _{\alpha }^{i_{\alpha }}(x_{\alpha } &\rightarrow &\infty )\propto
\exp (-k_{i_{\alpha }}x_{\alpha }) , \nonumber \\
\phi _{\alpha }^{i_{\alpha },out}(y_{\alpha } &\rightarrow &\infty )\propto
\exp (iq_{i_{\alpha }}y_{\alpha }) , \\
\Phi _{i_{\alpha }}^{out}(\rho &\rightarrow &\infty )\propto \exp (iK\rho ) ,
\label{eq:assf}
\end{eqnarray}
with $k_{i_{\alpha }}=\sqrt{-\varepsilon _{_{i_{\alpha }}}m}$ representing
momentum of 2-body bound state $i_{\alpha }$ with a negative binding energy $%
\varepsilon _{_{i_{\alpha }}}$; $q_{i_{\alpha }}=\sqrt{(E-\varepsilon
_{_{i_{\alpha }}})m}$ is relative scattering momentum for the $\alpha -(\beta
\gamma )$ binary channel, whereas $K=\sqrt{mE}$ is a three-particle breakup
momentum (three-particle breakup is possible only if energy value $E$ is
positive).

When considering particle's $\alpha $ scattering on the bound state $i_{\alpha
}$ of the pair $(\beta \gamma )$, it is
convenient to separate  readily
incoming wave $\psi _{\alpha }^{i_{\alpha },in}=\psi _{\alpha }^{i_{\alpha }}(\mathbf{x}_{\alpha })\phi _{\alpha }^{i_{\alpha },in}(\mathbf{y}_{\alpha
})$, by introducing:
\begin{eqnarray}
\psi _{\alpha }^{i_{\alpha },out} &=&\psi _{\alpha }^{i_{\alpha }}-\psi
_{\alpha }^{i_{\alpha }}(\mathbf{x}_{\alpha })\phi _{\alpha }^{i_{\alpha
},in}(\mathbf{y}_{\alpha }) , \\
\psi _{\beta }^{i_{\alpha },out} &=&\psi _{\beta }^{i_{\alpha }}\qquad \beta
\neq \alpha ,  \nonumber
\end{eqnarray}
Then Faddeev-Merkuriev equations might be rewritten in a so-called driven form:
\begin{eqnarray}
(E-H_{0}-\sum_{\kappa =1}^{3}w_{\kappa }^{l})\psi _{\alpha
}^{out}&=&(v_{\alpha }+w_{\alpha }^{s})(\psi _{\alpha }^{out}+\psi _{\beta
}^{out}+\psi _{\gamma }^{out})+\left[ \sum_{\kappa =1}^{3}w_{\kappa
}^{l}-w_{\alpha }-W_{\alpha }^{c.m.}\right] \psi _{\alpha }^{in} ,
 \nonumber \\
(E-H_{0}-\sum_{\kappa =1}^{3}w_{\kappa }^{l})\psi _{\beta }^{out}&=&(v_{\beta
}+w_{\beta }^{s})(\psi _{\alpha }^{out}+\psi _{\beta }^{out}+\psi _{\gamma
}^{out}+\psi _{\alpha }^{in}) , \\
(E-H_{0}-\sum_{\kappa =1}^{3}w_{\kappa }^{l})\psi _{\gamma
}^{out}&=&(v_{\gamma }+w_{\gamma }^{s})(\psi _{\alpha }^{out}+\psi _{\beta
}^{out}+\psi _{\gamma }^{out}+\psi _{\alpha }^{in}) . \nonumber
\label{eq:drive_FM}
\end{eqnarray}
In this expression index of the incoming state $i_{\alpha }$ has been omitted in
all Faddeev component expressions $\psi _{\alpha }^{in}$ and $\psi _{\alpha }^{out}$.
\subsection{Complex scaling}

Next step is to perform the complex scaling operations i.e. scale all the distances $%
x $ and $y$ by a constant complex factor $e^{i\theta },$ so that both $%
Re(e^{i\theta })$ and $Im(e^{i\theta })$ are positive (angle $\theta $ must
be chosen in the first quartet in order to satisfy this condition). The complex scaling
operation, in particular, implies that the analytical continuation of the
interaction potentials is performed: $v_{\alpha }(x_{\alpha }e^{i\theta })$
and $w_{\alpha }(x_{\alpha }e^{i\theta })$. Therefore the complex scaling method
may be used only if these potentials are analytic. It is easy to see that
the solutions of the complex scaled equations coincide with the ones obtained
without complex scaling but to which the complex scaling operation is
applied: $\left[ \psi (x_{\alpha },y_{\alpha })\right] ^{CS}=\psi (x_{\alpha
}e^{i\theta },y_{\alpha }e^{i\theta })$.

Namely, it is easy to demonstrate that all the outgoing wave functions of
eq.(\ref{eq:assf}) becomes exponentially bound after the complex scaling operation:
\begin{eqnarray}
\left[ \varphi _{\alpha }^{i_{\alpha }}(x_{\alpha }\rightarrow \infty )%
\right] ^{CS} &\propto &\exp (-k_{i_{\alpha }}x_{\alpha }\cos \theta ) , \nonumber\\
\left[ \phi _{\alpha }^{i_{\alpha },out}(y_{\alpha }\rightarrow \infty )%
\right] ^{CS} &\propto &\exp (-q_{i_{\alpha }}y_{\alpha }\sin \theta ) , \\
\left[ \Phi _{i_{\alpha }}^{out}(\rho \rightarrow \infty )\right] ^{CS}
&\propto &\exp (-K\rho \sin \theta ) . \nonumber
\end{eqnarray}

Nevertheless an incoming wave diverges in $y_{\alpha }$ after the complex
scaling:
\begin{equation}
\left[ \phi _{\alpha }^{i_{\alpha },out}(y_{\alpha }\rightarrow \infty )%
\right] ^{CS}\propto \exp (+q_{i_{\alpha }}y_{\alpha }\sin \theta ) .
\end{equation}
However these terms appear only on the right hand sides of the driven Faddeev-Merkuriev
 equation~(\ref{eq:drive_FM})
being pre-multiplied with the potential terms and under certain conditions they may vanish
outside of some finite (resolution) domain $x_{\alpha }\in \lbrack 0,x^{\max
}]$ and $y_{\alpha }\in \lbrack 0,y^{\max }]$. Let us consider the long range
behavior of the term $\left[ (v_{\beta }+w_{\beta }^{s})\psi _{\alpha }^{in}%
\right] ^{CS}$. Since the interaction terms $v_{\beta }$ and $w_{\beta }^{s}$
are of short range, the only region the former term might not converge is
along $y_{\beta }$ axis in $(x_{\beta },y_{\beta })$ plane, i.e. for $%
x_{\beta }\ll y_{\beta }$. On the other hand $x_{\alpha }(%
\mathbf{x}_{\beta }\mathbf{,y}_{\beta })\approx \sqrt{m_{\gamma }/(m_{\gamma
}+m_{\beta })}\sqrt{M/(m_{\gamma }+m_{\alpha })}y_{\beta }$ and $y_{\alpha }(%
\mathbf{x}_{\beta }\mathbf{,y}_{\beta })\approx \sqrt{m_{\beta }/(m_{\gamma
}+m_{\beta })}\sqrt{m_{\alpha }/(m_{\gamma }+m_{\alpha })}y_{\beta }$ under
condition $x_{\beta }\ll y_{\beta }$. Then one has:
\begin{equation}
\small{
\left[ (v_{\beta }+w_{\beta }^{s})\psi _{\alpha }^{i_{\alpha },in}\right]
^{CS}_{x_{\beta }\ll y_{\beta }}\propto \exp\left(-k_{i_{\alpha }}\sqrt{\frac{%
m_{\gamma }M}{(m_{\gamma }+m_{\beta })(m_{\gamma }+m_{\alpha })}}y_{\beta
}\cos \theta +q_{i_{\alpha }}\sqrt{\frac{m_{\alpha }m_{\beta }}{%
(m_{\gamma }+m_{\beta })(m_{\gamma }+m_{\alpha })}}y_{\beta }\sin \theta \right)} .
\end{equation}
This term becomes bound to finite domain in $(x_{\beta },y_{\beta }) $
plane, if condition:
\begin{equation}
\tan \theta <\sqrt{\frac{m_{\gamma }M}{m_{\alpha }m_{\beta }}}\frac{%
k_{i_{\alpha }}}{q_{i_{\alpha }}}=\sqrt{\frac{m_{\gamma }M}{m_{\alpha
}m_{\beta }}}\sqrt{\frac{\left\vert B_{_{i_{\alpha }}}\right\vert }{%
E+\left\vert B_{_{i_{\alpha }}}\right\vert }} ,
\label{max_theta}
\end{equation}
is satisfied. This implies that for rather large scattering energies $E$,
above the break-up threshold, one is obliged to use rather small complex
scaling parameter $\theta $ values.

The term $\left[\sum_{\kappa =1}^{3}w_{\kappa }^{l}-w_{\alpha }-W_{\alpha
}^{c.m.}\right] \psi _{\alpha }^{i_{\alpha },in}$, in principle, is not
exponentially bound after the complex scaling. It represents the higher
order corrections to the residual Coulomb interaction between particle $%
\alpha $ and bound pair $(\beta \gamma )$. These corrections are weak $%
o(1/y^{2})$ and might be neglected by suppressing this term close to the
border of the resolution domain. Alternative possibility might be to use
incoming wave functions, which account not only for the bare $\alpha -(\beta
\gamma )$ Coulomb interaction but also takes into account higher order
polarization corrections.

\bigskip

Extraction of the scattering observables is realized by employing Greens
theorem. One might demonstrate that strong interaction amplitude for $%
\alpha -(\beta \gamma )$ collision is:
\begin{equation}
f_{j_{\alpha i_{\kappa }}}(\mathbf{x}_{\alpha }.\mathbf{y}_{\alpha })=-\frac{m}{q_{j_{\alpha}}}\int
\int \left[ (\psi _{\alpha }^{j_{\alpha },in})^*\right] ^{CS}(\overline{v}%
_{\alpha }+\overline{w}_{\alpha }-W_{\alpha }^{c.m.})^{CS}\left[ \Psi
_{i_{\kappa }}\right] ^{CS}e^{6i\theta }d^{3}\mathbf{x}_{i}d^{3}\mathbf{y}%
_{i}  \label{3b_amp_nc} ,
\end{equation}%
with $\left[ \Psi _{i_{\kappa }}\right] ^{CS}=\left[ \psi _{\alpha
}^{i_{\kappa },out}+\psi _{\beta }^{i_{\kappa },out}+\psi _{\gamma
}^{i_{\kappa },out}+\psi _{\alpha }^{i_{\kappa },in}\right] ^{CS}$ being the
total wave function of the three-body system. In the last expression the
term containing product of two incoming waves is slowest to converge. Even
stronger constraint than eq.(\ref{max_theta}) should be implied on complex scaling angle in
order to make this term integrable on the finite domain. Nevertheless this term
contains only the product of two-body wave functions and might be evaluated
without using complex scaling prior to three-body solution. Then the appropriate
form of the integral~(\ref{3b_amp_nc}) to be used becomes:
\begin{eqnarray}
f_{j_{\alpha i_{\kappa }}}(\mathbf{x}_{\alpha }.\mathbf{y}_{\alpha })
&=&-\frac{m}{q_{j_{\alpha}}}\int \int \left[ (\psi _{\alpha }^{j_{\alpha },in})^*\right] ^{CS}(%
\overline{v}_{\alpha }+\overline{w}_{\alpha }-W_{\alpha }^{c.m.})^{CS}\left[
\Psi _{i_{\kappa }}-\psi _{\alpha }^{j_{\alpha },in}\right] ^{CS}e^{6i\theta
}d^{3}\mathbf{x}_{i}d^{3}\mathbf{y}_{i} \nonumber\\
&&-\frac{m}{q_{j_{\alpha}}}\int \int (\psi _{\alpha }^{j_{\alpha },in})^*(\overline{v}_{\alpha }+%
\overline{w}_{\alpha }-W_{\alpha }^{c.m.})\psi _{\alpha }^{j_{\alpha
},in}d^{3}\mathbf{x}_{i}d^{3}\mathbf{y} .
\end{eqnarray}

\bigskip

\bigskip

\section{Application to three-body nuclear reactions}

The two methods presented in sections~\ref{sec:p} and~\ref{sec:r} were first
applied to the proton-deuteron elastic scattering and
breakup~\cite{deltuva:05a,deltuva:05d,deltuva:09e,lazauskas:11a}.
The three-nucleon system is the
only nuclear three-particle system that may be considered realistic in the sense that the interactions
are given by high precision potentials valid over a broad energy range.
Nevertheless, in the same way one considers the nucleon as a single particle by neglecting
its inner quark structure, in a further approximation one can
consider a cluster of nucleons (composite nucleus)
to be a single particle that interacts with other nucleons or nuclei via
effective potentials whose parameters are determined from the two-body data.
 A classical example is the $\alpha$ particle, a tightly
bound four-nucleon cluster. As shown in Figs.~\ref{fig:Rad} and \ref{fig:Radb}
and in Ref.~\cite{deltuva:06b}, the description of the
 $(\alpha,p,n)$ three-particle system with real potentials
is quite successful at low
energies but becomes less reliable with increasing energy where the
inner structure of the $\alpha$ particle cannot be neglected anymore.
At higher energies the nucleon-nucleus or nucleus-nucleus interactions
are modeled by optical potentials (OP) that provide quite an accurate
description of the considered two-body system
in a given narrow energy range; these potentials are complex to account for the inelastic
excitations not explicitly included in the model space. The  methods based on
Faddeev/AGS equations can be applied also in this case, however,
the potentials within the pairs that are bound in the initial or final
channel must remain real. The comparison of the two methods based
on the AGS and FM equations  will be performed in section~\ref{sec:compare}
 for such an interaction model with OP.

In the past the description of three-body-like nuclear reactions involved
a number of approximate methods that have been developed. Well-known examples are
the distorted-wave Born approximation (DWBA), various adiabatic approaches
\cite{johnson:70a}, and
continuum-discretized coupled-channels (CDCC) method~\cite{austern:87}.
Compared to them the present methods based on exact Faddeev or AGS equations,
being  more technically and numerically involved,
have some disadvantages. Namely,
their application in the present technical realization is so far limited
to a system made of two nucleons and one heavier cluster.
The reason is
that the interaction between two  heavier cluster involves
very many angular momentum states and the partial-wave convergence
cannot be achieved. The comparison between traditional nuclear
reaction approaches and momentum-space Faddeev/AGS methods
for various neutron + proton + nucleus systems
are summarized in section~\ref{sec:cdcc}.

On the other hand, the  Faddeev and AGS  methods may be more flexible
with respect to dynamic input and thereby allows to test novel aspects
of the nuclear interaction not accessible with the traditional approaches.
Few examples will be presented in section \ref{sec:nonloc}.

\subsection{Numerical comparison of AGS and FM methods} \label{sec:compare}

As an example we consider the $n+p+^{12}C$ system. For the $n$-$p$ interaction
we use a realistic AV18 model~\cite{wiringa:95a} that accurately reproduces the
available two-nucleon scattering data and deuteron binding energy.
To study not only the $d+{}^{12}$C but also  $p+{}^{13}$C
scattering and transfer reactions we use
a $n$-$^{12}$C potential that is real in the $^2P_\frac{1}{2}$
partial wave and supports the ground state of $^{13}C$
with 4.946 MeV binding energy; the parameters are taken from
Ref.~\cite{nunes:11b}. In all other partial waves we
use the $n$-$^{12}$C optical  potential from Ref.~\cite{CH89}
taken at half the deuteron energy in the $d+{}^{12}$C channel.
The $p$-$^{12}$C optical potential is also taken from Ref.~\cite{CH89}, however, at the proton energy in the $p+{}^{13}$C channel.
We admit that, depending on the reaction of interest,
other choices of energies for OP may be more appropriate,
however, the aim of the present study is comparison of the
methods and not the description  of the experimental data although the latter
are also included in the plots.

\begin{figure}
\sidecaption[t]
\includegraphics[scale=.56]{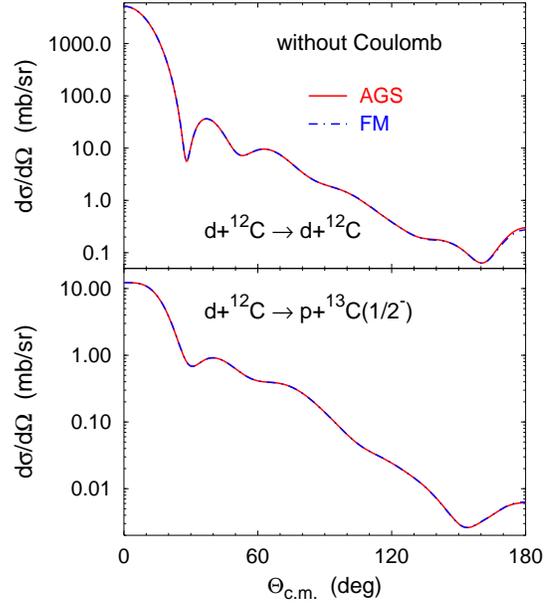}
\caption{
Comparison of momentum- (solid curves) and configuration-space
(dashed-dotted curves) results for the
deuteron-${}^{12}$C scattering at 30 MeV deuteron lab energy.
Differential cross sections for elastic scattering and stripping
are shown neglecting the Coulomb interaction.}
\label{fig:dC-noC}
\end{figure}

We consider $d+{}^{12}$C scattering at 30 MeV deuteron lab energy
and  $p+{}^{13}$C scattering at 30.6 MeV proton lab energy;
they correspond to the same energy in c.m. system.
First we perform calculations by neglecting the $p$-$^{12}$C Coulomb repulsion.
One observes a perfect agreement between the AGS and FM methods.
 Indeed, the calculated S-matrix elements in each three-particle
channel considered (calculations have been performed for total three-particle
angular momentum states up to $J=13$) agree  within
 three digits. Scattering observables  converge quite slowly with $J$
as different angular momentum state contributions cancel each other
at large angles.
Nevertheless, the results of the two methods are practically
indistinguishable as demonstrated in Fig.~\ref{fig:dC-noC}
for $d+{}^{12}$C elastic scattering and transfer to $p+{}^{13}$C.

\begin{figure}
\sidecaption[t]
\includegraphics[scale=.56]{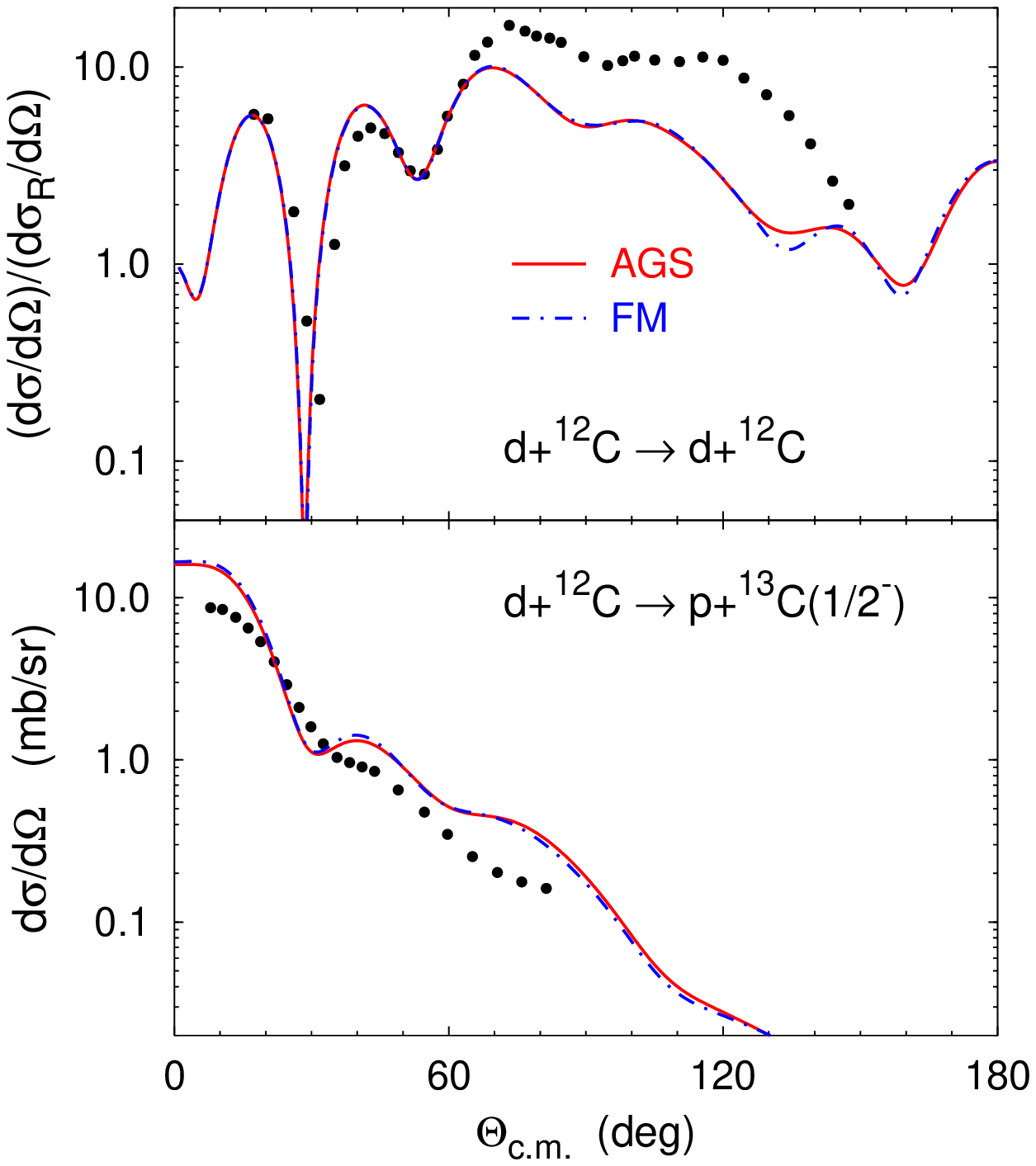}
\caption{
Comparison of momentum- (solid curves) and configuration-space
(dashed-dotted curves) results for the
deuteron-${}^{12}$C scattering at 30 MeV deuteron lab energy.
Differential cross sections for elastic scattering and stripping
are shown, the former in ratio to the Rutherford cross section
$d\sigma_R/d\Omega$.
The experimental data are from Refs.~\cite{perrin:77,dC30p}.}
\label{fig:dC}
\end{figure}

\begin{figure}
\sidecaption[t]
\includegraphics[scale=.56]{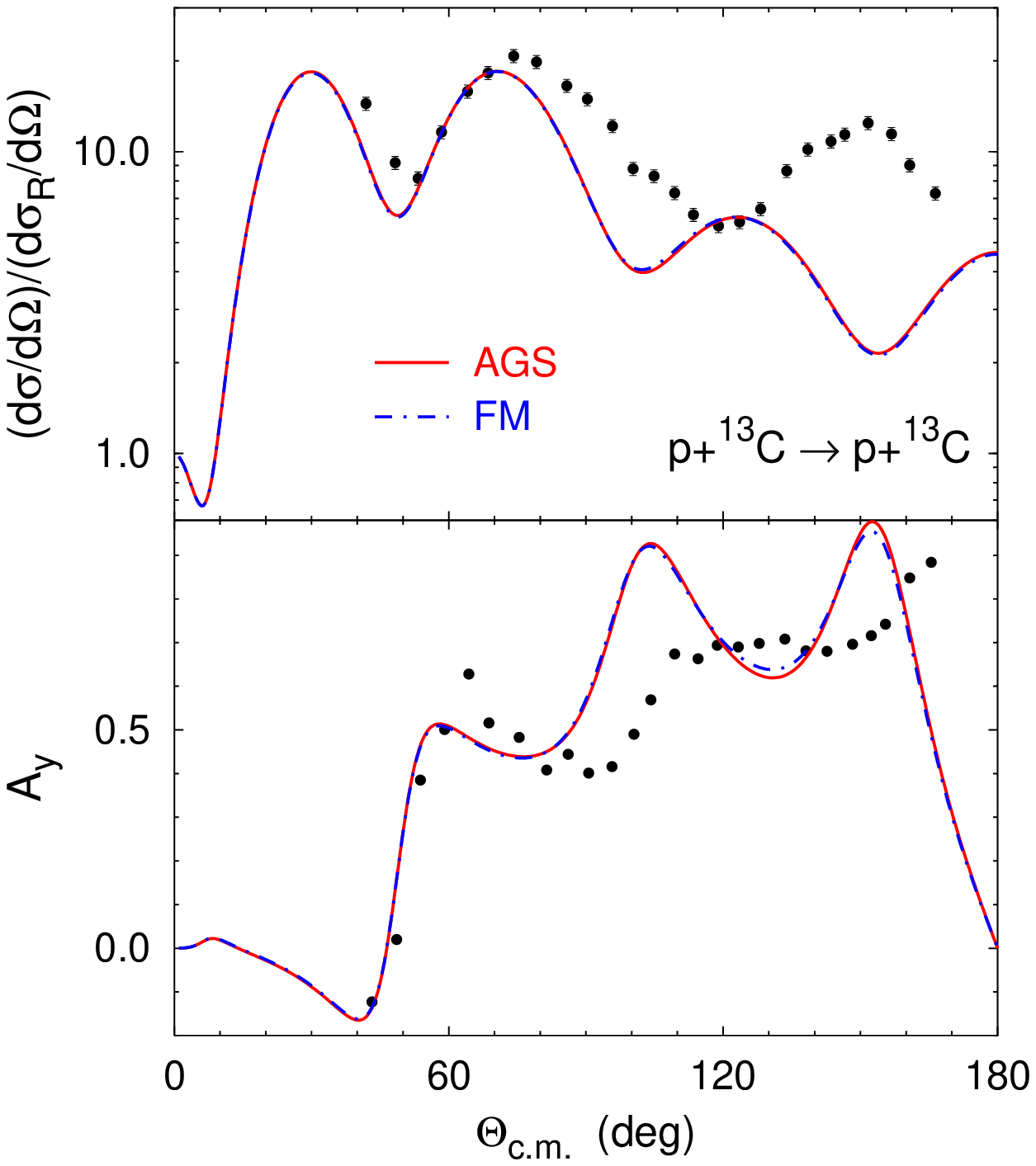}
\caption{
Comparison of momentum- (solid curves) and configuration-space
(dashed-dotted curves) results for the
proton-${}^{13}$C elastic scattering at 30.6 MeV proton lab energy.
Differential cross section divided by the Rutherford cross section
and proton analyzing power are shown.
The experimental data are from Ref.~\cite{pC30}.}
\label{fig:pC}
\end{figure}

Next we perform the full calculation including the $p$-$^{12}$C
Coulomb repulsion;  we note that inside the nucleus the Coulomb potential
is taken as the one of a uniformly charged sphere~\cite{deltuva:06b}.
Once again we obtain good agreement between the AGS and FM methods.
However, this time small variations up to the order of 1\% are observed when
analyzing separate $S$-matrix elements, mostly in high angular momentum states.
 This leads to small differences in some scattering observables, e.g.,
differential cross sections for $d+{}^{12}$C elastic scattering
(at large angles where the differential cross section is very small)
and for the deuteron stripping reaction $d+{}^{12}$C $ \to p+{}^{13}$C
shown in Fig.~\ref{fig:dC}. The  $p+{}^{13}$C  elastic scattering observables
presented in Fig.~\ref{fig:pC} converge faster with $J$.
As a consequence, the results of the two calculations are indistinguishable
 for the $p+{}^{13}$C elastic cross section
and only tiny differences  can be seen for the proton analyzing power
at large angles. In any case, the agreement between the AGS and FM methods
exceeds both the accuracy of the data and the existing discrepancies
between theoretical predictions and experimental data.

\subsection{Comparison with traditional nuclear reaction approaches} \label{sec:cdcc}

\begin{figure}
\includegraphics[scale=.45]{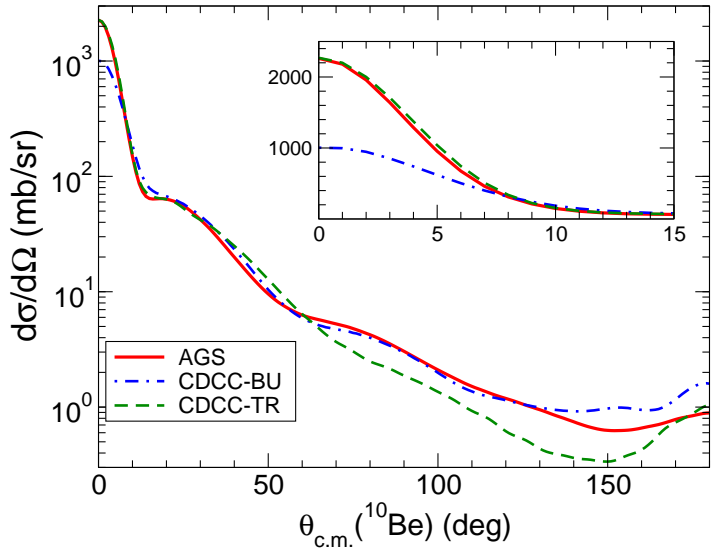}
\caption{Semi-inclusive differential cross section for the
breakup reaction $p + {}^{11}$Be $\to p + n + {}^{10}$Be
at lab energy of 38.4 MeV/nucleon. Results obtained with
AGS and CDCC methods are compared.}
\label{fig:cdcc}
\end{figure}

The method based on the momentum-space AGS equations has already been
used to test the accuracy of the traditional nuclear reaction approaches;
 limitations of their validity in energy and kinematic range have
been estalished.
The distorted-wave impulse approximation  for breakup of a one-neutron
halo nucleus ${}^{11}$Be on a proton target has been tested in
Ref.~\cite{crespo:08a}
while the adiabatic-wave approximation for the deuteron stripping
and pickup reactions ${}^{11}$Be$(p,d){}^{10}$Be,
${}^{12}$C$(d,p){}^{13}$C, and ${}^{48}$Ca$(d,p){}^{49}$Ca
in Ref.~\cite{nunes:11b}. However,
one of the most sophisticated traditional  approaches
is the CDCC method~\cite{austern:87}. A detailed comparison between
CDCC and AGS results is performed in Ref.~\cite{deltuva:07d}.
The agreement is good for deuteron-${}^{12}$C and deuteron-${}^{58}$Ni
elastic scattering and breakup. In these cases nucleon-nucleus interactions
were given by optical potentials; thus, there was no transfer reaction.
A different situation takes place in proton-${}^{11}$Be scattering
where ${}^{11}$Be nucleus is assumed to be the bound state of a
${}^{10}$Be core plus a  neutron. In this case, where the transfer
channel $d + {}^{10}$Be is open, the CDCC approach lacks
accuracy as shown in Ref.~\cite{deltuva:07d}.
The semi-inclusive differential cross section for the
breakup reaction $p + {}^{11}$Be $\to p + n + {}^{10}$Be
was calculated also using two CDCC versions where the full scattering wave function
was expanded into the eigenstates of either the $n + {}^{10}$Be (CDCC-BU)
or the $p+n$ (CDCC-TR) pair. Neither of them agrees well with AGS
over the whole angular regime as shown in Fig.~\ref{fig:cdcc}.
It turns out that, depending on the
${}^{10}$Be scattering angle, the semi-inclusive breakup cross section
is dominated by different mechanisms: at small angles it is the
proton-neutron quasifree scattering whereas at intermediate and large angles
it is the neutron-${}^{10}$Be $D$-wave resonance. However, a proper treatment
of proton-neutron interaction in CDCC-BU and of  neutron-${}^{10}$Be
interaction in CDCC-TR  is very hard to achieve since the wave function
expansion uses eigenstates of a different pair.
No such problem exists in the AGS method that uses simultaneously
three sets of basis states and each pair is treated in its proper basis.

\subsection{Beyond standard dynamic models} \label{sec:nonloc}

The standard nucleon-nucleus optical potentials employed
in three-body calculations have central and, eventually, spin-orbit parts
that are local.
This local approximation yields a tremendous simplification in the
practical realization of DWBA, CDCC and other traditional approaches
that are based on configuration-space representations where
 the use of nonlocal optical potentials was never attempted.
However, nonlocal optical potentials do not yield any serious technical
difficulties in the momentum-space representation. Thus, they
can be included quite easily in the AGS framework employed by us.

There are very few nonlocal parametrizations of the optical potentials
available. We take the one from Refs.~\cite{giannini,giannini2} defined
in the  configuration space as
\begin{equation}  \label{eq:vnl}
v_{\gamma}(\vec{r}',\vec{r}) = H_c(x)[V_c(y) + iW_c(y)] +
2\vec{S_\gamma}\cdot \vec{L_\gamma} H_s(x) V_s(y) ,
\end{equation}
with $x = |\vec{r}'-\vec{r}|$ and  $y=|\vec{r}'+\vec{r}|/2$.
The central part has real volume and imaginary surface parts,
whereas the  spin-orbit part is real; all of them are expressed in the standard
way by Woods-Saxon functions. Some of their strength parameters were
 readjusted in Ref.~\cite{deltuva:09b}
to improve the description of the experimental nucleon-nucleus scattering
data. The range of the nonlocality is determined by the functions
$H_i(x) = (\pi \beta_i^2)^{-3/2} \exp{(-x^2/\beta_i^2)}$
with the parameters $\beta_i$ being of the order of 1 fm.

\begin{figure}
\sidecaption[t]
\includegraphics[scale=.56]{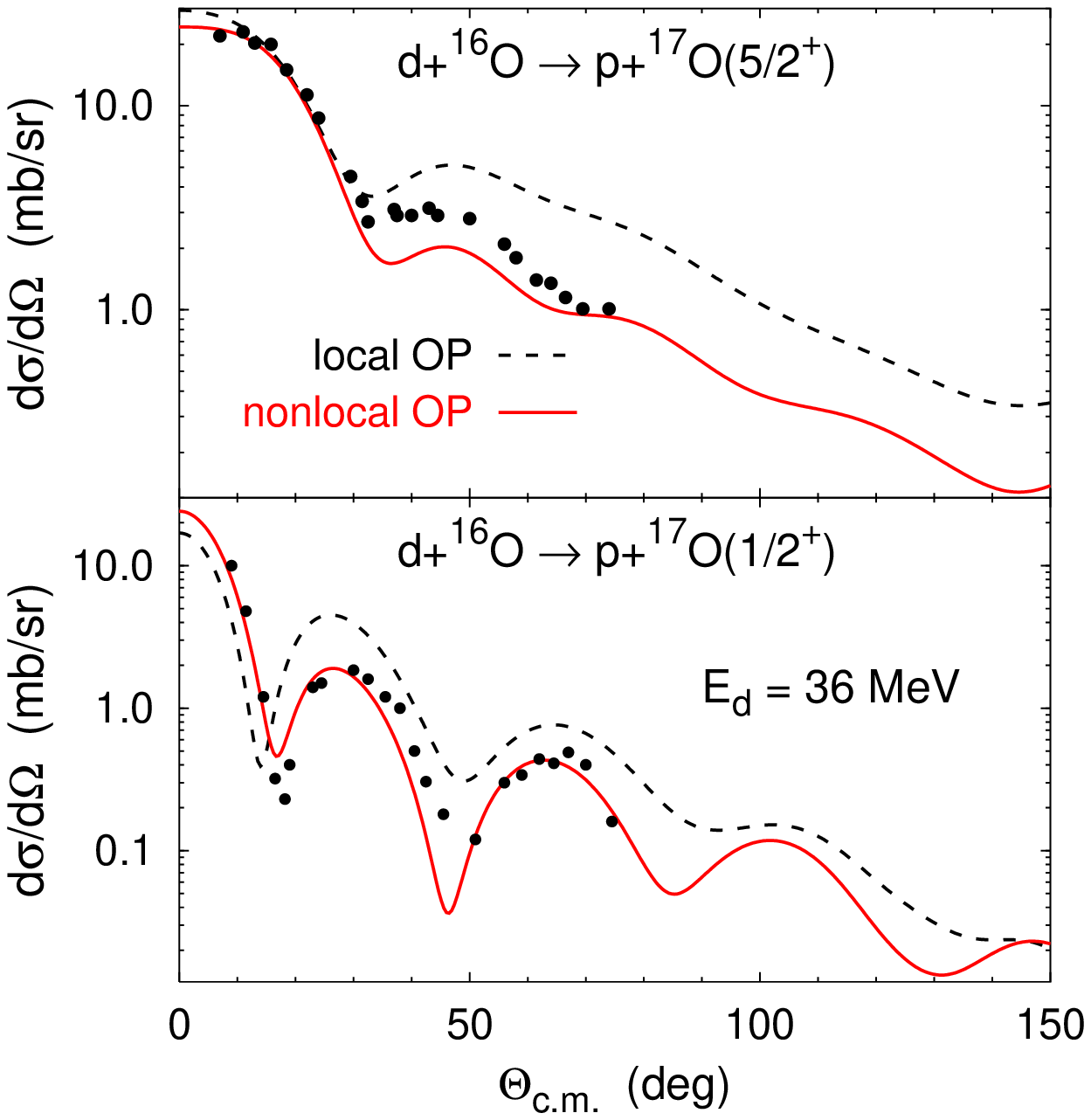}
\caption{
Differential cross section for $(d,p)$ reaction on ${}^{16}$O
at 36 MeV deuteron lab energy
leading to ${}^{17}$O nucleus in the ground state $5/2^+$ (top)
and first excited state $1/2^+$ (bottom).
Predictions of nonlocal (solid curve) and local
(dashed curve) optical potentials (OP)
are compared with the
 experimental data from Ref.~\cite{dO25-63}.}
\label{fig:Odp}
\end{figure}

\begin{figure}
\sidecaption[t]
\includegraphics[scale=.56]{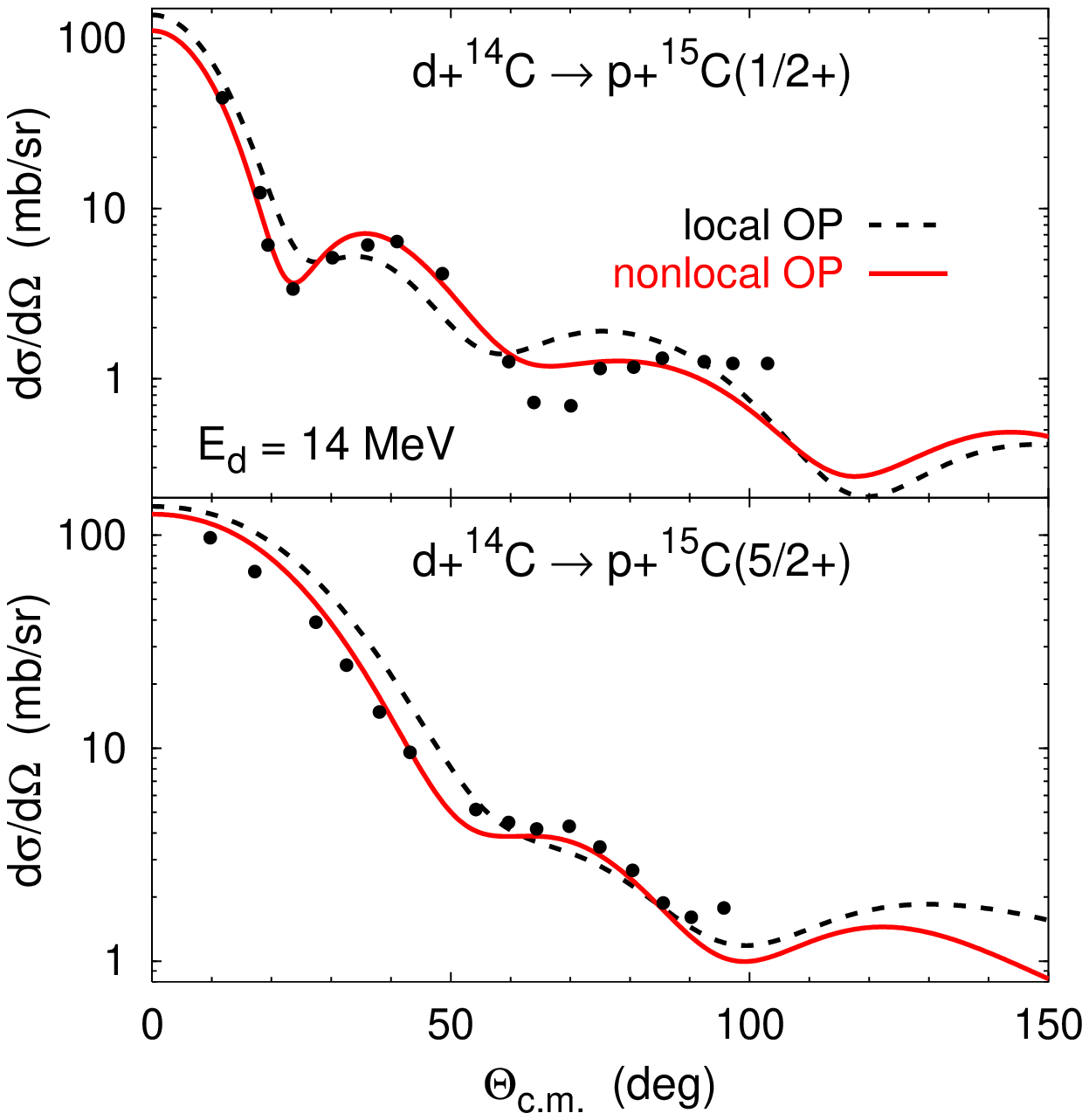}
\caption{
Differential cross section for $(d,p)$ reaction on ${}^{14}$C
at 14 MeV deuteron lab energy
leading to one-neutron halo nucleus
${}^{15}$C  in the ground state $1/2^+$ (top)
and first excited state $5/2^+$ (bottom).
Curves as in Fig.~\ref{fig:Odp} and the
 experimental data are from Ref.~\cite{d14C14p}.}
\label{fig:Cdp}
\end{figure}

A detailed study of nonlocal optical potentials in three-body reactions
involving stable as well as weakly bound nuclei, ranging from
${}^{10}$Be to ${}^{40}$Ca,
is carried out in Ref.~\cite{deltuva:09b}.
In order to isolate the nonlocality effect
we also performed calculations with a local optical potential that
provides approximately equivalent description
of the nucleon-nucleus scattering at the considered energy.
The nonlocality effect turns out to be very small in the elastic proton
scattering from the bound neutron-nucleus system and of moderate
size in the deuteron-nucleus scattering. However, the
effect of  nonlocal proton-nucleus optical  potential becomes
significant in deuteron stripping and pickup reactions $(d,p)$ and $(p,d)$;
in most cases it considerably improves agreement with the experimental data.
Examples for $(d,p)$ reactions leading to ground and excited states of
the stable nucleus ${}^{17}$O and one-neutron halo nucleus ${}^{15}$C
are presented in Figs.~\ref{fig:Odp} and \ref{fig:Cdp}.
We note that in these transfer reactions the proton-nucleus potential
is taken at proton lab energy in the proton channel while
the neutron-nucleus potential has to be real in order to support the
respective  bound states.

Another extension beyond the standard dynamic models includes
the AGS method using energy-dependent optical potentials
Although such calculations don't correspond
to a rigorous Hamiltonian theory, they may shed some light on the
shortcomings of the traditional nuclear interaction models.
A detailed discussion of the calculations with
energy-dependent optical potentials is given in Ref.~\cite{deltuva:09a}.

\section{Summary}

We have presented the results of three-body Faddeev-type calculations for systems of three  particles, two of which are charged, interacting through short-range nuclear plus the long-range Coulomb potentials. Realistic applications of three-body theory to three-cluster nuclear reactions --- such as  scattering of deuterons on a nuclear target or one-neutron halo nucleus impinging on a proton target ---  only became possible to address in recent years
when a reliable and practical  momentum-space treatment of the Coulomb
interaction has been developed.
After the extensive and very complete study of $p$-$d$ elastic scattering and breakup, the natural extension of these calculations was the application to complex reactions such as $d$-${}^{4}$He, $p$-${}^{17}$O, ${}^{11}$Be-$p$, $d$-${}^{58}$Ni and many others using a realistic interaction such as AV18 between nucleons, and optical potentials chosen at the appropriate energy for the nucleon-nucleus interactions. The advantage of three-body calculations vis-\`{a}-vis traditional approximate reaction methods  is that elastic, transfer, and breakup channels are treated on the same footing once the interaction Hamiltonian has been chosen. Another advantage of the three-body Faddeev-AGS approach is the possibility to include
nonlocal optical potentials instead of local ones as commonly used in the standard nuclear reaction methods; as demonstrated, this leads to an improvement in the description of transfer reactions in a very consistent way across different energies and mass numbers for the core nucleus.

Although most three-body calculations have been performed in momentum space over a broad range of nuclei from ${}^{4}$He to ${}^{58}$Ni and have encompassed studies of cross sections and polarizations for elastic, transfer, charge exchange,
 and breakup reactions, coordinate space calculations above breakup threshold are coming to age using the complex scaling method. We have demonstrated here that both calculations agree to within a few percent for all the reactions we have calculated. This is a very promising development that may bring new light to the study of nuclear reactions given that the reduction of the many-body problem to an effective three-body one may be better implemented and understood by the community in coordinate space rather than in momentum space. On the other hand,
  compared to DWBA, adiabatic approaches, or CDCC, the
Faddeev-type three-body methods are computationally more demanding and require greater technical expertise rendering them less attractive to analyze the data.
Nevertheless, when benchmark calculations have been performed comparing the Faddeev-AGS results with those obtained using CDCC or adiabatic approaches, some discrepancies were found in transfer and breakup cross sections depending on the specific kinematic conditions. Therefore the Faddeev-AGS approach is imminent in order to calibrate and validate approximate nuclear reaction methods wherever a comparison is possible.

\begin{acknowledgement}
The work of A.D. and A.C.F. was partially supported by the
FCT grant PTDC/FIS/65736/2006.
The work of R.L. was granted access to the HPC resources of IDRIS under the allocation 2009-i2009056006
made by GENCI (Grand Equipement National de Calcul Intensif). We thank the staff members of the IDRIS for their constant help.
\end{acknowledgement}


\end{document}